\documentclass[twocolumn,english,aps,prb,twocolum,superscriptaddress,bibnotes,amsmath,amssymb,floatfix]{revtex4-1}
\usepackage[colorlinks=true,citecolor=blue,linkcolor=magenta]{hyperref}

\usepackage[utf8]{inputenc}
\usepackage[english]{babel}
\usepackage{amsmath,amsfonts,amssymb}
\usepackage[T1]{fontenc}
\usepackage{url}

\usepackage{amsmath}
\usepackage{amsfonts}
\usepackage{amssymb}
\usepackage[version=3]{mhchem}

\usepackage{epstopdf}
\usepackage{graphicx}
\graphicspath{{./Figures/}}

\begin{document}

\title{Ultralow-power chip-based soliton microcombs for photonic integration}

\author{Junqiu Liu}
\thanks{J.L. and A.S.R contributed equally to this work}
\affiliation{{\'E}cole Polytechnique F{\'e}d{\'e}rale de Lausanne (EPFL), CH-1015 Lausanne,
Switzerland}
 
\author{Arslan S. Raja}
\thanks{J.L. and A.S.R contributed equally to this work}
\affiliation{{\'E}cole Polytechnique F{\'e}d{\'e}rale de Lausanne (EPFL), CH-1015 Lausanne,
Switzerland}

\author{Maxim Karpov}
\affiliation{{\'E}cole Polytechnique F{\'e}d{\'e}rale de Lausanne (EPFL), CH-1015 Lausanne,
Switzerland}

\author{Bahareh Ghadiani}
\affiliation{{\'E}cole Polytechnique F{\'e}d{\'e}rale de Lausanne (EPFL), CH-1015 Lausanne,
Switzerland}

\author{Martin H. P. Pfeiffer}
\affiliation{{\'E}cole Polytechnique F{\'e}d{\'e}rale de Lausanne (EPFL), CH-1015 Lausanne,
Switzerland}

\author{Nils J. Engelsen}
\affiliation{{\'E}cole Polytechnique F{\'e}d{\'e}rale de Lausanne (EPFL), CH-1015 Lausanne,
Switzerland}

\author{Hairun Guo}
\affiliation{{\'E}cole Polytechnique F{\'e}d{\'e}rale de Lausanne (EPFL), CH-1015 Lausanne,
Switzerland}

\author{Michael Zervas}
\affiliation{LIGENTEC SA, CH-1015 Lausanne, Switzerland}

\author{Tobias J. Kippenberg}
\email[]{tobias.kippenberg@epfl.ch}
\affiliation{{\'E}cole Polytechnique F{\'e}d{\'e}rale de Lausanne (EPFL), CH-1015 Lausanne,
Switzerland}

\date{\today}
\maketitle

\noindent\textbf{\noindent
The generation of dissipative Kerr solitons in optical microresonators has provided a route to compact frequency combs of high repetition rate, which have already been employed for optical frequency synthesizers, ultrafast ranging, coherent telecommunication and dual-comb spectroscopy. Silicon nitride (Si$_3$N$_4$) microresonators are promising for photonic integrated soliton microcombs. Yet to date, soliton formation in Si$_3$N$_4$ microresonators at electronically detectable repetition rates, typically less than 100 GHz, is hindered by the requirement of external power amplifiers, due to the low quality ($Q$) factors, as well as by thermal effects which necessitate the use of frequency agile lasers to access the soliton state. These requirements complicate future photonic integration, heterogeneous or hybrid, of soliton microcomb devices based on Si$_3$N$_4$ microresonators with other active or passive components. Here, using the photonic Damascene reflow process, we demonstrate ultralow-power single soliton formation in high-Q ($Q_0>15\times10^6$) Si$_3$N$_4$ microresonators with 9.8 mW input power (6.2 mW in the waveguide) for devices of electronically detectable, 99 GHz repetition rate. We show that solitons can be accessed via simple, slow laser piezo tuning, in many resonances in the same sample. These power levels are compatible with current silicon-photonics-based lasers for full photonic integration of soliton microcombs, at repetition rates suitable for applications such as ultrafast ranging and coherent communication. Our results show the technological readiness of Si$_3$N$_4$ optical waveguides for future all-on-chip soliton microcomb devices.
}

\def \DWRep {\Delta \omega_{\rm rep}}
\def \DFRep {\Delta f_{\rm rep}}


\section{Introduction}

Optical frequency combs \cite{Udem:02, Cundiff:03} have revolutionized timekeeping and metrology over the past decades, and have found a wide variety of applications \cite{Newbury:11}. First discovered more than a decade ago, microresonator-based Kerr frequency combs ("microcomb") \cite{DelHaye:07, Kippenberg:11} are providing a route to chip-scale optical frequency combs, with broad bandwidth and repetition rates in the GHz to THz domain. Such microcombs have been demonstrated in a wide variety of platforms, including CMOS compatible materials such as silicon nitride ($\mathrm{Si_3N_4}) $ \cite{Levy:10, Moss:13}, which allows full photonic integration with other devices on silicon, active or passive, such as lasers \cite{Liang:10}, modulators \cite{Reed:10} and photodetectors \cite{Michel:10}. The demonstration of microcombs in the dissipative Kerr soliton (DKS) regime \cite{HerrNP:14} has unlocked the full potential of microcombs, by allowing reliable access to fully coherent comb states, which can attain large bandwidth via soliton Cherenkov radiation \cite{Jang:14, Milian:14, Brasch:15}. Such soliton microcombs have already been successfully applied in counting the cycles of light \cite{Jost:15}, coherent communication on the receiver and transmitter sides \cite{Marin-Palomo:2017}, dual-comb spectroscopy \cite{Suh:16}, ultrafast optical ranging \cite{Trocha:17, Suh:17}, astrophysical spectrometer calibration \cite{Obrzud:17, Suh:18}, and for creating a photonic integrated frequency synthesizer \cite{Spencer:17}. These developments highlight the potential of chip-scale soliton microcombs for timing, metrology and spectroscopy, allowing unprecedentedly compact devices at low operation power, fully compatible with wafer-scale fabrication and suitable for operation in space \cite{Brasch:14}. 

A particularly promising platform for photonic integrated soliton microcombs is Si$_3$N$_4$, a material that has a wide transparency window and high material optical nonlinearity. Recent advances in fabrication have enabled access to the anomalous group velocity dispersion regime (GVD) with sufficient waveguide height \cite{Foster:06}, and circumvented the problems associated with the high tensile stress of Si$_3$N$_4$ film \cite{Luke:13}. Although the soliton microcomb has been directly generated with a diode laser in a silica microresonator coupled to a tapered optical fiber recently \cite{Volet:18}, this has not been possible for integrated devices including Si$_3$N$_4$. Yet, there are remaining challenges in soliton formation in Si$_3$N$_4$ microresonators, related to: (i) The comparatively low quality ($Q$) factor which increases the threshold power of soliton formation, compared to e.g. crystalline and silica microresonators. (ii) Optical coupling losses from the laser to the chip device, resulting from the optical mode mismatch at chip facets. (iii) Stable access to soliton states may require the use of complex excitation techniques such as "power kicking" \citep{Brasch:16} or single-sideband modulators \cite{Stone:18}. The first two challenges are particularly problematic for future photonic integration, as they necessitate the use of high-power lasers. So far, for integrated Si$_3$N$_4$ microresonator devices, soliton formation with device input power of several hundreds of milliwatts has only been achieved in microresonators of 1 THz free spectral range (FSR) \citep{Li:17, Pfeiffer:17, Spencer:17, Briles:18}. Yet in these experiments input coupling loss still necessitated the use of additional amplifiers to reach the required power levels of several tens of milliwatts on the chip. Meanwhile, low microcomb initiation power at milliwatt or even sub-milliwatt level has been demonstrated in Si$_3$N$_4$ microresonators of $Q$ exceeding $10^7$ \cite{Xuan:16, Ji:17}, but solitons have not been observed due to the insufficient anomalous GVD. Consequently, soliton formation has been limited to wavelength regions where optical amplifiers are available. It is only very recently that soliton generation in high-Q Si$_3$N$_4$ microresonators of anomalous GVD pumped by an integrated laser was reported in Ref. \cite{Stern:18}. However in that case, the repetition rate is 200 GHz, which is not electronically detectable, resulting in limited application potentials.


\begin{figure}[t!]
\centering
\includegraphics{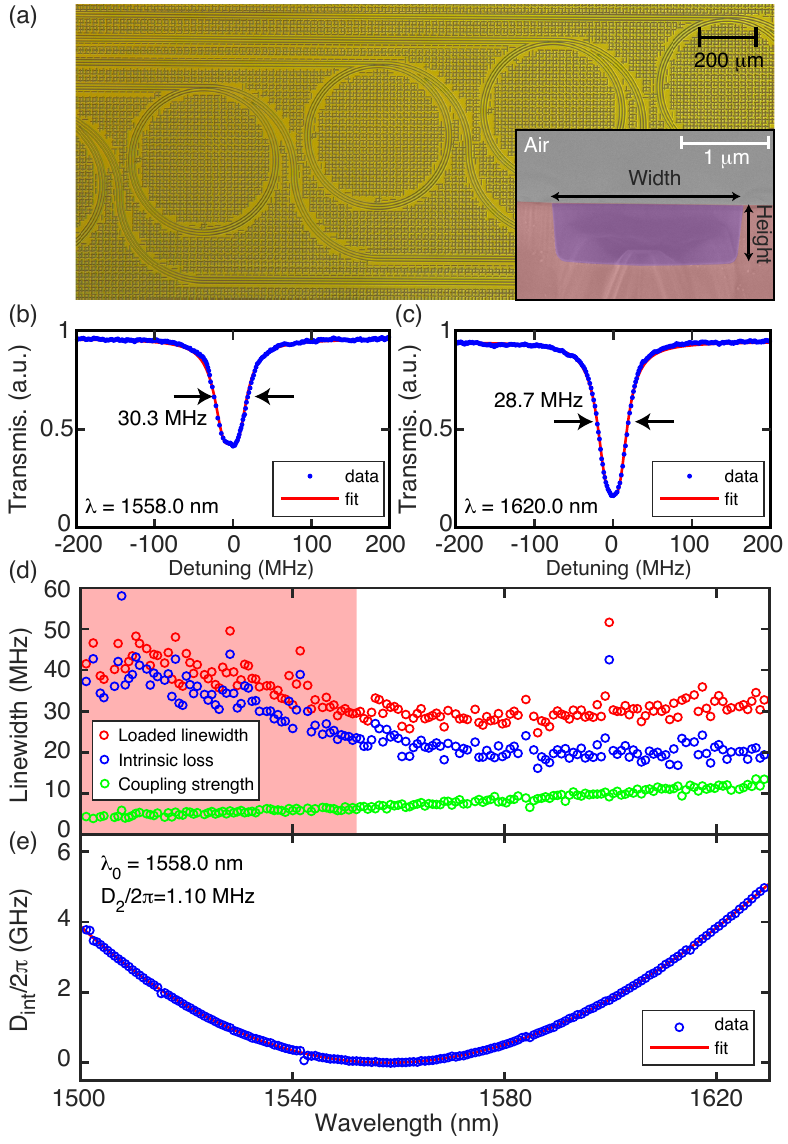}
\caption{Dispersion and resonance linewidth characterization of a 88-GHz-FSR microresonator. (a) Microscope image of densely packed 88-GHz-FSR microresonators, using meander bus waveguides. These samples have no SiO$_2$ top cladding. Inset: SEM image of the cross-section of a Si$_3$N$_4$ waveguide without SiO$_2$ top cladding. The Si$_3$N$_4$ waveguide is blue shaded, the SiO$_2$ bottom cladding is red shaded, the air is not color shaded. (b) The pump resonance at $\lambda=1558.0$ nm and its fit, with the loaded linewidth of $\kappa/2\pi=30.3$ MHz and fitted intrinsic loss of $\kappa_0/2\pi\sim 23.2$ MHz, corresponding to $Q_0>8.2\times10^6$. (c) The resonance at $\lambda=1620.0$ nm and its fit, with the loaded linewidth of $\kappa/2\pi=28.7$ MHz and fitted intrinsic loss of $\kappa_0/2\pi\sim 17.3$ MHz, corresponding to $Q_0>10.7\times10^6$. (d) Loaded linewidth, intrinsic loss and coupling strength of each TE$_{00}$ resonance. Larger intrinsic loss is found in the wavelength region from 1500 to 1550 nm (red shaded area), due to the absorption by the N-H and Si-H bonds in LPCVD Si$_3$N$_4$. (e) Measured GVD of the TE$_{00}$ mode family. Several avoided mode crossings are observed, where resonance linewidth increases.}
\label{Fig:figure4}
\end{figure}

Here we demonstrate that the newly developed variant of photonic Damascene process \cite{Pfeiffer:16, Pfeiffer:18b}, the \emph{photonic Damascene reflow} process \cite{Pfeiffer:18}, can overcome the outlined challenges and significantly reduce the required input power for soliton formation in Si$_3$N$_4$ microresonators. We demonstrate single soliton formation in Si$_3$N$_4$ microresonators of $Q_0>8.2\times10^6$ at the lowest repetition rate to date of 88 GHz, which is electronically detectable, with 48.6 mW power at the chip input facet (30.3 mW in the bus waveguide). In addition, by further improving the microresonator Q factors to $Q_0>15\times10^6$, we demonstrate single soliton formation of 99 GHz repetition rate with a record-low input power of 9.8 mW (6.2 mW in the waveguide). Using only a tunable diode laser without an optical amplifier, we access single soliton states in eleven consecutive resonances in the telecom L-band and five in the telecom C-band, via simple laser piezo tuning. Such low-power soliton microcombs of sub-100-GHz repetition rate can significantly simplify the recently demonstrated dual-comb ultrafast distance measurements \cite{Trocha:17} and optical coherent communication \cite{Marin-Palomo:2017}, which required Erbium-doped fiber amplifiers (EDFA) to amplify the input power to above 1 Watt. In addtion, the soliton microcombs demonstrated here have shown great potential for future photonic integrated microwave generators, and chip-based frequency synthesizers \cite{Spencer:17} via integration of on-chip lasers, semiconductor optical amplifiers and nonlinear microresonators. Soliton microcombs formed in wavelength regions where amplifiers are not available could unlock new applications such as optical coherent tomography (OCT) at 1.3 $\mu$m \cite{Yun:03} and sensing of toxic gases and greenhouse gases e.g. methane at 1.6 $\mu$m \cite{Tombez:17}. 

\section{Sample fabrication}

Here, we briefly describe our fabrication process of Si$_3$N$_4$ microresonator samples. The waveguide is patterned on photoresist on the silicon dioxide (SiO$_2$) substrate, using deep-UV (DUV) stepper lithography. The pattern is then transferred from the DUV photoresist to the SiO$_2$ substrate via reactive ion etching (RIE) using C$_4$F$_8$, O$_2$ and helium. Before the Si$_3$N$_4$ deposition on the patterned substrate using low pressure chemical vapor deposition (LPCVD), we do a "preform reflow" \cite{Pfeiffer:18}, where the substrate is annealed at 1250$^\circ$C temperature with atmospheric pressure. Due to the high temperature and the SiO$_2$ surface tension, the sidewall roughness caused by the RIE on the SiO$_2$ preform is reduced, which reduces scattering losses and leads to ultra-smooth waveguide side surfaces. This helps to improve the $Q$ of our Si$_3$N$_4$ microresonators.

The LPCVD Si$_3$N$_4$ film is deposited on the SiO$_2$ substrate, filling the preform trenches and defining the Si$_3$N$_4$ waveguides. The deposition is followed by chemical-mechanical polishing (CMP) which removes the excess Si$_3$N$_4$ and creates an ultrasmooth waveguide top surface. The substrate is annealed at 1200$^\circ$C in nitrogen atmosphere, to drive out the residual hydrogen content introduced from the SiH$_2$Cl$_2$ and NH$_3$ utilized in the LPCVD Si$_3$N$_4$ process. Adding SiO$_2$ top cladding is optional, while both situations are considered in our work, which will be described later. Finally, the wafer is separated into chips by dicing or with deep RIE. More details of fabrication process are found in Ref. \cite{Pfeiffer:18b} and \cite{Pfeiffer:18}.

For integrated chip-based nonlinear photonics, inverse nanotapers \citep{Almeida:03} are widely used as chip input couplers and can achieve high fiber-chip-fiber coupling efficiency and broad operation bandwidth. The standard subtractive process \cite{Gondarenko:09, Xuan:16} requires the taper width to be less than 100 nm, to achieve fiber-chip-fiber coupling efficiency of 40$\%$ or more. Thus the resolution of DUV stepper lithography is incompatible with the substrative process, which requires instead electron beam lithography. For the Damascene process, the required optimum taper width is above 400 nm, to attain 40$\%$ coupling efficiency, thus the DUV stepper lithography is compatible with this process. The high coupling efficiency with much larger taper width, compared with the one used in the subtractive process, is due the strong aspect-ratio-dependent etch rate \cite{Gottscho:92} of the preform RIE, which is specifically engineered to allow the creation of \emph{double-inverse nanotapers} as chip input couplers (more details are found in Ref. \cite{Liu:18}). The double-inverse nanotaper enables high coupling efficiency (40$\%$ or more) with large taper width (above 400 nm) for both the transverse-electric (TE) and transverse-magnetic (TM) polarizations, due to the reduced taper height, whcih increases the evanescent field size at the taper tip, thus improving the mode match between the taper mode to the incident lensed fiber mode \cite{Liu:18}.

\begin{figure}[t!]
\centering
\includegraphics{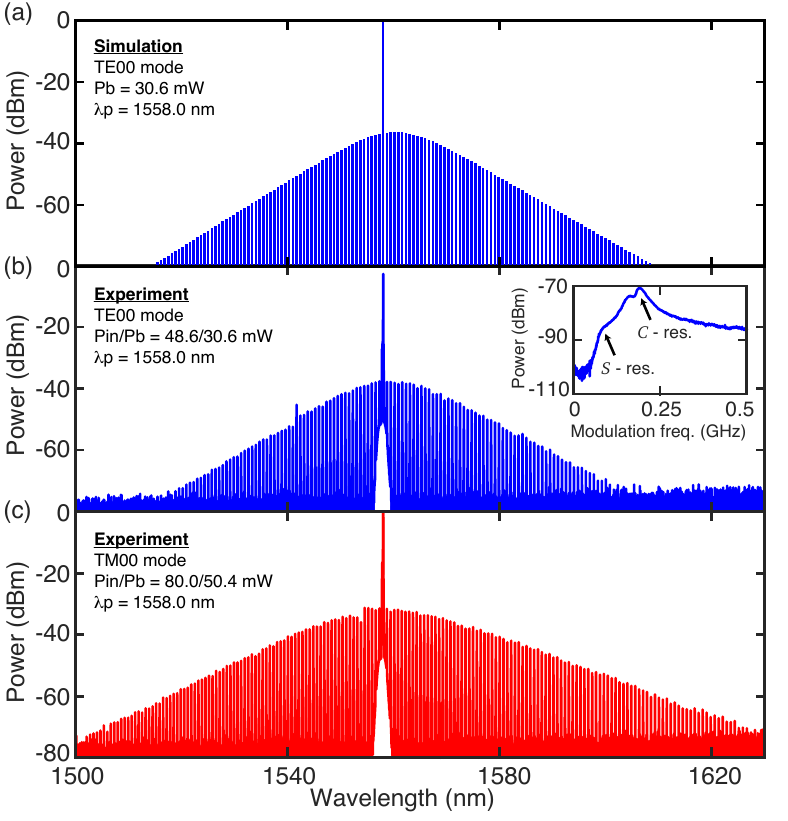}
\caption{Single soliton formation in the 88-GHz-FSR microresonator. (a) Simulated single soliton spectrum based on the measured microresonator's parameters, in the TE$_{00}$ mode. $P_\text{b}=30.6$ mW corresponds to $P_\text{in}=48.6$ mW. (b) Single soliton spectrum pumped at $\lambda_\text{p}=1558.0$ nm in the TE$_{00}$ mode, with a pump power of $P_\text{in}=48.6$ mW ($P_\text{b}=30.6$ mW). Inset: cavity response measurement using the VNA, verifying that the spectrum is a single soliton state. (c) Single soliton spectrum pumped at $\lambda_\text{p}=1558.0$ nm in the TM$_{00}$ mode, with the pump power $P_\text{in}=80.0$ mW ($P_\text{b}=50.4$ mW).}
\label{Fig:figure5}
\end{figure}

\section{Soliton comb of 88 GHz repetition rate}

In the first part, we demonstrate single soliton formation in 88-GHz-FSR microresonators, with the fundamental TE mode (TE$_{00}$). Fig. \ref{Fig:figure4}(a) shows the microscope image of the 88-GHz-FSR microresonators. The microresonator samples described in this section have no SiO$_2$ top cladding, as shown in Fig. \ref{Fig:figure4}(a) inset. Meander bus waveguides are used to densely pack a large number of devices on one chip. The Si$_3$N$_4$ microresonator has a cross-section, width$\times$height, of $1.58 \times 0.75$  $\mu$m$^2$, and is coupled to a multi-mode bus waveguide of the same cross-section for high coupling ideality \cite{Pfeiffer:17b}. The polarization of the incident light to the chip is controlled by fiber polarization controllers, and the polarization state is measured using linear polarizers. The microresonator transmission trace is obtained from 1500 to 1630 nm using frequency-comb-assisted diode laser spectroscopy \cite{Delhaye:09, Liu:16}. The precise frequency of each data point is calibrated using a commercial femtosecond optical frequency comb with 250 MHz repetition rate. For the TE$_{00}$ mode family, the FSR of the microresonator and the anomalous GVD are extracted from the calibrated transmission trace by identifying the precise frequency of each resonance. The total (loaded) linewidth $\kappa/2\pi=(\kappa_0+\kappa_\text{ex})/2\pi$, the intrinsic linewidth (intrinsic loss) $\kappa_0/2\pi$ and the coupling strength $\kappa_\text{ex}/2\pi$ are extracted from each resonance fit \cite{Gorodetsky:00, Li:13}.

The measured linewidths of each TE$_{00}$ resonance are shown in Fig. \ref{Fig:figure4}(d). These resonances are all undercoupled. Larger linewidths are found in the wavelength region from 1500 to 1550 nm, due the absorption by the nitrogen-hydrogen (N-H) and silicon-hydrogen (Si-H) bonds in LPCVD Si$_3$N$_4$ material \cite{Mao:08, Bauters:11}. These bonds are introduced during standard LPCVD Si$_3$N$_4$ process based on SiH$_2$Cl$_2$ and NH$_3$ precursors \cite{Yota:00}, and can be partially removed by thermal annealing. Fig. \ref{Fig:figure4}(b) shows the resonance at $\lambda=1558.0$ nm. The resonance fit shows a loaded linewidth of $\kappa/2\pi=30.3$ MHz, and the estimated intrinsic loss based on the resonance fit is $\kappa_0/2\pi\sim 23.3$ MHz, corresponding to the intrinsic $Q_0>8.2\times10^6$. Fig. \ref{Fig:figure4}(c) shows the resonance at $\lambda=1620.0$ nm, with a loaded linewidth of $\kappa/2\pi=28.7$ MHz, and the estimated intrinsic loss is $\kappa_0/2\pi\sim 17.3$ MHz, corresponding to the intrinsic $Q_0>10.7\times10^6$. Fig. \ref{Fig:figure4}(e) shows the measured microresonator integrated GVD, defined as $D_\text{int}(\mu)=\omega_{\mu}-\omega_0-D_1\mu=D_2\mu^2/2+D_3\mu^3/6+...$ \cite{Liu:16}. Here $\omega_{\mu}/2\pi$ is the frequency of the $\mu$-th resonance relative to the reference resonance $\omega_0/2\pi=192.6$ THz ($\lambda_0=1558.0$ nm, as shown in Fig.\ref{Fig:figure4}(b)). $D_1/2\pi$ corresponds to the FSR. $D_2$, $D_3$ and other higher order terms determine the GVD profile. The fit values obtained from the dispersion measurement are $D_1/2\pi\sim88.63$ GHz, $D_2/2\pi\sim1.10$ MHz and $D_3/2\pi\sim\mathcal{O}(1)$ kHz.

\begin{figure}[t!]
\centering
\includegraphics{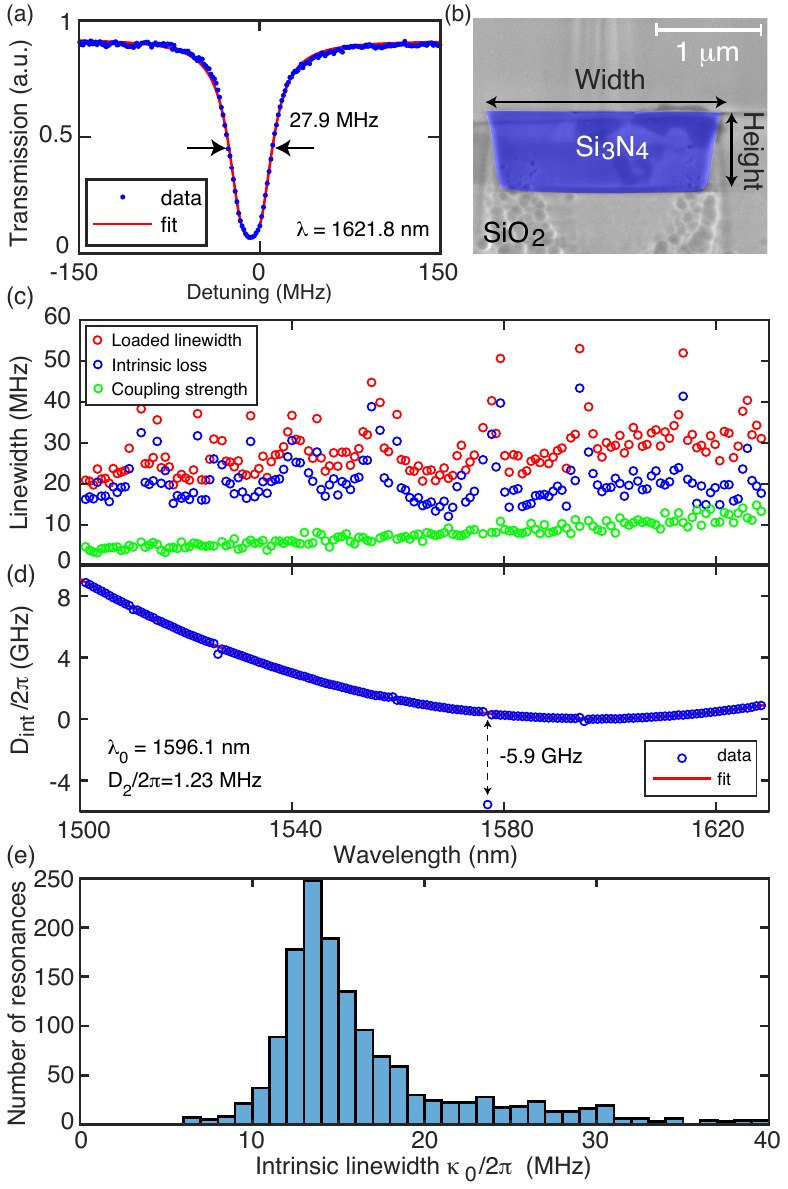}
\caption{Dispersion and resonance linewidth characterization of a 99-GHz-FSR microresonator. (a) The critically coupled resonance at $\lambda=1621.8$ nm and its fit, with the loaded linewidth of $\kappa/2\pi=27.9$ MHz and fitted intrinsic loss of $\kappa_0/2\pi\sim 13.7$ MHz. (b) SEM image of the cross-section of a Si$_3$N$_4$ waveguide with full SiO$_2$ cladding. The Si$_3$N$_4$ waveguide is blue shaded, the SiO$_2$ cladding is not color shaded. (c) Loaded linewidth, intrinsic loss and coupling strength of each TE$_{00}$ resonance. No prominent hydrogen absorption loss is observed in the wavelength region from 1500 to 1550 nm. (d) Measured GVD of the TE$_{00}$ mode family. Several avoided mode crossings are observed, where resonance linewidth increases. A strong mode crossing is found at 1577 nm, with -5.9 GHz resonance frequency deviation. (e) Histogram of intrinsic loss from the measurement of eight under-coupled samples. The most probable value of the histogram is around 13 - 14 MHz, which represents the Q factor of $Q_0>15\times10^6$. }
\label{Fig:figure6}
\end{figure}

When the pump laser scans over the resonance from the blue-detuned side to the red-detuned side, a step in the transmission trace can be observed, signaling to the soliton formation \cite{HerrNP:14}. To generate and characterize soliton states, we use a setup as describe in Ref. \cite{Guo:16}. Fiber-chip-fiber transmission of $40\%$ (coupling efficiency of $63\%$ per device facet) is achieved via double-inverse nanotapers on the chip facets \cite{Liu:18}. The input power ($P_\text{in}$) is defined as the power measured before the input lensed fiber which couples light into the chip device. Thus the optical power in the bus waveguide ($P_\text{b}$) on the chip, which directly pumps the microresonator, is calculated as $P_\text{b}=0.63P_\text{in}$.

In our setup, the output power of the diode laser can go as high as 23 mW, with few milliwatt power variation depending on the wavelength. We use an EDFA to slightly amplify the optical power to $P_\text{in}=48.6$ mW ($P_\text{b}=30.6$ mW). To access the single soliton state, we use a single-sideband modulator \cite{Stone:18}, and a single soliton spectrum is observed as shown in Fig. \ref{Fig:figure5}(b), verified by the system response measurement using vector network analyzer (VNA) \cite{Guo:16}. The observed double resonance response shown in Fig. \ref{Fig:figure5}(b) inset corresponds to the cavity resonance of the continuous wave ("$\mathcal{C}$-resonance"), and the soliton-induced resonance ("$\mathcal{S}$-resonance"). These two resonances can be distinguished by increasing the detuning of the soliton state \citep{Guo:16}. Fig. \ref{Fig:figure5}(a) shows the simulation of soliton formation based on Lugiato-Lefever equation \cite{Lugiato:87, Coen:13} using the measured microresonator parameters. The simulated soliton spectrum is nearly identical to the measured one. We also generate a single soliton in the TM$_{00}$ mode ($D_1/2\pi\sim86.35$ GHz, $D_2/2\pi\sim0.967$ MHz, $D_3/2\pi\sim5.4$ kHz) pumped at the wavelength $\lambda_\text{p}=1558.0$ nm, with the input power $P_\text{in}=80.0$ mW ($P_\text{b}=50.6$ mW), as shown in Fig. \ref{Fig:figure5}(c). This soliton spectrum is broader than the one in the TE$_{00}$ mode shown in Fig. \ref{Fig:figure5}(b), likely due to the lower $D_2$ value and higher power. The estimated power conversion efficiency from the CW pump to the soliton pulse is around 1.5$\%$, with the power per comb line around 20 $\mu$W in the 3 dB bandwidth. Compared to prior works shown in Ref. \cite{Marin-Palomo:2017, Trocha:17} which use solitons of repetition rate less than 100 GHz with input power exceeding 1 Watt, our work represents a significant power reduction, while the power per comb line of 20 $\mu$W still can achieve the same goals as Ref.  \cite{Marin-Palomo:2017, Trocha:17}.

\section{Soliton comb of 99 GHz repetition rate}

\begin{figure}[t!]
\centering
\includegraphics{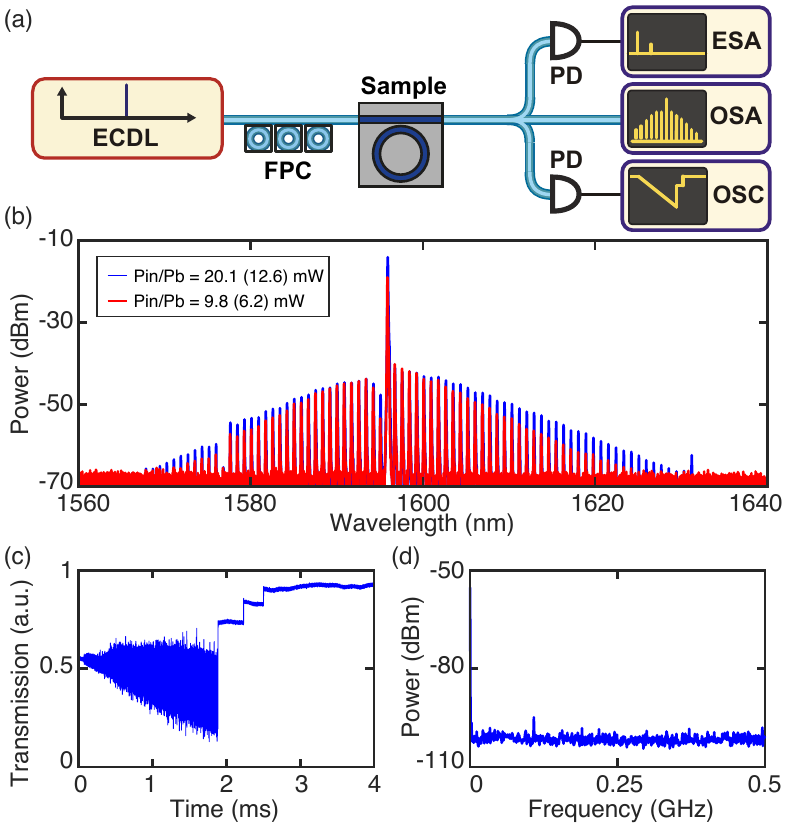}
\caption{Single soliton formation in a 99-GHz-FSR microresonator without EDFA. (a) Experimental setup. ECDL: external-cavity diode laser. OSC: oscilloscope. OSA: optical spectrum analyzer. ESA: electrical spectrum analyzer. FPC: fiber polarization controller. PD: photodiode. (b) Single soliton spectra pumped at $\lambda_\text{p}=1596.1$ nm in the TE$_{00}$ mode, with the input pump powers of $P_\text{in}=9.8$ mW ($P_\text{b}=6.2$ mW, red) and $P_\text{in}=20.1$ mW ($P_\text{b}=12.6$ mW, blue). (c) A representative soliton step of several hundred of microsecond, sufficiently long for accessing the single soliton state via simple laser piezo tuning. (d) Low-frequency RF spectrum of the optical spectrum with $P_\text{in}=9.8$ mW (red), demonstrating the soliton nature of the spectrum.}
\label{Fig:figure7}
\end{figure}

In the second part of our work, we demonstrate single soliton formation in 99-GHz-FSR microresonators. Fig. \ref{Fig:figure4}(d) shows that hydrogen absorption is the main loss reason which prevents the generation of 88 GHz soliton with lower power. Therefore we further improved the fabrication process, in order to remove hydrogen absorption losses. Note that, the hydrogen is likely introduced due to the incomplete thermal annealing, or moisture in the air which forms a thin water film on the sample surface, or a combination of both two. Therefore in a new wafer fabrication run, we annealed the LPCVD Si$_3$N$_4$ via deposition - annealing - deposition - annealing cycles, as described in Ref. \citep{Luke:15}. To prevent water film formation on the wafer, a thick SiO$_2$ top cladding composed of TEOS and low temperature oxide (LTO) was deposited via LPCVD on the Si$_3$N$_4$ waveguides (SEM image of the waveguide cross-section is shown in Fig. \ref{Fig:figure6}(b)), followed by thermal annealing. Simultaneously, the dry etching process which patterns the SiO$_2$ preform was optimized, to further reduce mask damage and sidewall passivation, which can reduce the waveguide sidewall roughness and the number of defects. Fig. \ref{Fig:figure6}(c) shows the loaded linewidth, intrinsic loss and coupling strength of each TE$_{00}$ resonance of a Si$_3$N$_4$ microresonator whose cross-section, width$\times$height, is $1.58 \times 0.81$  $\mu$m$^2$. The microresonator FSR is $D_1/2\pi\sim98.9$ GHz. Compared with Fig. \ref{Fig:figure4}(d), Fig. \ref{Fig:figure6}(c) shows significant reduction of intrinsic loss in the wavelength range from 1500 to 1550 nm, demonstrating the successful removal of hydrogen in LPCVD Si$_3$N$_4$. Some resonances with large linewidth are caused by avoided mode crossings, in accordance with the observed avoided mode crossings in the dispersion measurement shown in Fig. \ref{Fig:figure6}(d). The measured GVD is $D_2/2\pi\sim1.23$ MHz, with respect to $\omega_0/2\pi=188.0$ THz ($\lambda_0=1596.1$ nm as the pump resonance in Fig. \ref{Fig:figure7}). Figure \ref{Fig:figure6}(e) shows the histogram of intrinsic loss from the measurement of eight under-coupled samples. These samples have the same waveguide cross-section but different bus-waveguide-to-microresonator gap distance. The most probable value of the histogram is around 13 - 14 MHz, which represents the Q factor of $Q_0>15\times10^6$. Fig. \ref{Fig:figure6}(a) shows the resonance at $\lambda=1621.8$ nm and its fit, with the loaded linewidth of $\kappa/2\pi=27.9$ MHz and fitted intrinsic loss of $\kappa_0/2\pi\sim 13.7$ MHz.

\begin{figure*}[t!]
\centering
\includegraphics{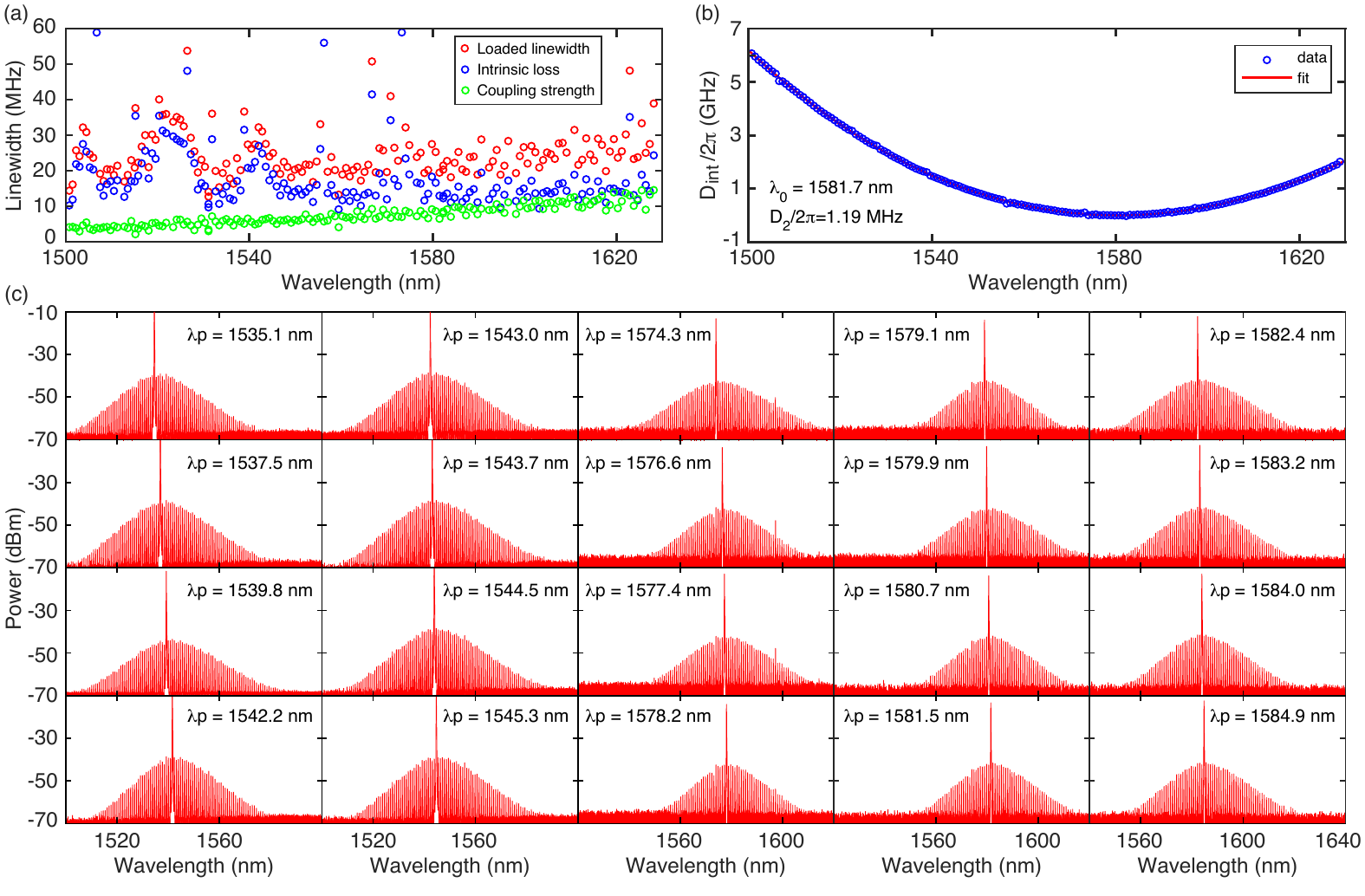}
\caption{Characterization of a different 99-GHz-FSR microresonator, and single soliton formation in multiple resonances. (a) Loaded linewidth, intrinsic loss and coupling strength of each TE$_{00}$ resonance. (b) Measured GVD of the TE$_{00}$ mode family. (c) Single soliton formation in twenty selected resonances, eleven of which are consecutive in the telecom L-band, and five of which are consecutive in the telecom C-band. $\lambda_\text{p}$ is the wavelength of the pumped resonance. The complete hydrogen removal facilitates the soliton generation in the hydrogen absorption band.}
\label{Fig:figure8}
\end{figure*}

Silicon nitride microresonators of anomalous GVD and Q factor exceeding $10\times10^6$ have been reported \cite{Xuan:16, Ji:17}, however single soliton generation has not been demonstrated. In those works, large waveguide width ($\geq 2.5$ $\mu$m) was used. Despite the fact that the large waveguide width reduces optical mode interaction with the waveguide sidewall roughness, and thus reduces the scattering loss caused by the sidewall roughness, the resulted weak anormalous GVD due to the large waveguide width is insufficient for single soliton generation with low power. Here, with the high Q and strong anomalous GVD ($D_2/2\pi\sim1.23$ MHz), we generate a single soliton of 99 GHz repetition rate with 9.8 mW input power (6.2 mW power in the bus waveguide), directly from the diode laser, without EDFA. The experimental setup is shown in Fig. \ref{Fig:figure7}(a), and the single soliton spectra are shown in Fig. \ref{Fig:figure7}(b). Parametric oscillation which generates frequency sidebands is observed around 1.7 mW chip input power. When the diode laser scans over the resonance, the observed soliton step varies from several hundred of microsecond to a millisecond (a representative soliton step in the transmision trace is shown in Fig. \ref{Fig:figure7}(c)), which is sufficiently long for accessing the single soliton state via simple laser piezo tuning \cite{HerrNP:14}. Increasing power to the maximum laser output (around 20.1 mW) increases the soliton bandwidth. The estimated power conversion efficiency from the CW pump to the soliton pulse is around 1.7$\%$. To identify the soliton nature of the spectrum, in this case a VNA measurement is difficult to implement due to the large EOM insertion loss and the limited diode laser output power. Instead, the soliton nature is revealed by the low-frequency radio frequency (RF) spectrum, as shown in Fig. \ref{Fig:figure7}(d), which can be well distinguished from the recorded noisy comb spectrum (modulation instability, not shown here). We observed that, the single soliton generation with less than 10 mW input power is highly reproducible in resonances close to avoided mode crossings. In our case, without the EDFA and its gain bandwidth limitation, we tune the diode laser frequency to a resonance which is close to a mode crossing, and investigate the minimum power for single soliton generation. We have experimentally measured several samples, and observed that such sub-10-mW-power single soliton generation is highly reproducible in these resonances, all of which feature long soliton steps (similar to the one shown in Fig. \ref{Fig:figure7}(c)). It appears that the mode crossings can facilitate single soliton formation with lower power, compared with normal resonances far from mode crossings. This phenomenon is likely due to single soliton generation assisted by spatial mode-interactions \cite{Bao:17}.

We further investigate the single soliton formation over the full tuning range of the diode laser. Fig. \ref{Fig:figure8}(a), (b) show the measured resonance linewidth and microresonator dispersion of a different 99-GHz-FSR microresonator ($D_1/2\pi\sim98.9$ GHz, $D_2/2\pi\sim1.19$ MHz, $D_3/2\pi\sim-1.1$ kHz). Fig. \ref{Fig:figure8}(c) shows the single soliton spectra pumped at twenty selected resonances, eleven of which are consecutive in the telecom L-band, and five of which are consecutive in the telecom C-band. These spectra are generated with laser output power around 22 mW, and all accessed via simple laser piezo tuning. The complete hydrogen removal facilitates the soliton generation in the hydrogen absorption band. We did not investigate the minimum power to generate soliton in these resonances. We have also accessed single or few soliton state in other resonances in the same sample, in addition to the ones shown in Fig. \ref{Fig:figure8}(c). The single soliton generation in a broad range of resonances demonstrates the reliability of our fabrication process, and offers extra flexibility to investigate spectrally localized effects, such as avoided mode crossings, which can enable the formation of dark pulses in the normal GVD region \cite{Liu:14, Xue:15}, and breathing solitons \cite{Guo:17}. 

\section{Conclusion}

In summary, we present ultralow-power single soliton formation in integrated high-$Q$ Si$_3$N$_4$ microresonators of sub-100-GHz FSR. We demonstrate soliton formation in a 88-GHz-FSR microresonator at 48.6 mW input power. The ultralow-power soliton formation results mainly from the microresonator's high $Q$ (estimated intrinsic $Q_0>8.2\times10^6$) and the >40$\%$ device-through coupling efficiency. In addition, by further improving the microresonator Q factor to $Q_0>15\times10^6$ and reducing the thermal effects due to hydrogen absorption, we demonstrate single soliton formation of 99 GHz repetition rate with a record-low input power of 9.8 mW (6.2 mW in the waveguide), accessed via simple laser piezo tuning. We demonstrate the simplicity of soliton tuning and result reproducibility in twenty selected resonances in the same sample. Our work demonstrates soliton microcomb generation in Si$_3$N$_4$ integrated microresonators with milliwatts power level, central for applications which require low power consumption, such as photonic chip-based microwave generators, integrated frequency synthesizers, OCT and dual-comb spectroscopy. 

\emph{Data availability}: The code and data used to produce the plots within this Article are available on Zenodo \cite{Zenodo}. All other data used in this study are available from the corresponding authors upon reasonable request.

\section*{Funding Information}
This work was supported by Contract HR0011-15-C-0055 (DODOS) from the Defense Advanced Research Projects Agency (DARPA), Defense Sciences Office (DSO), and by Swiss National Science Foundation under grant agreement No. 161573 (precoR).


\section*{Acknowledgments}
The Si$_3$N$_4$ microresonator samples were fabricated in the EPFL center of MicroNanoTechnology (CMi). M.K. acknowledges the support from the European Space Technology Centre with ESA Contract No. 4000116145/16/NL/MH/GM and 4000118777/16/NL/GM respectively. H.G. acknowledges the support from the European Union’s Horizon 2020 research and innovation program under Marie Sklodowska-Curie IF grant agreement No. 709249. We acknowledge Tiago Morais for the assistance in sample fabrication.


\bibliographystyle{apsrev4-1}
\bibliography{Bibcolec}
\end{document}